
\baselineskip 15pt
\magnification=1200

\def\d {\rm d}

\def\sk#1{${\cal S} #1$}
\def\tft {topological field theory }
\def\braket#1 { \langle \, #1  \, \rangle }
\def\Braket#1 {$\braket#1 $}
\def\ket#1 {$ | \, #1 \,  \rangle$ }
\def\bra#1 {$ \langle \, #1  \, | $ }
\def\inner#1#2 { (  #1  \, , \, #2 ) }
\def\Dim#1 {{\rm Dim } \, #1 }

\def\spnet {spin network }
\def\qsymm#1 {{\quad \Bigl {|} \quad \over \quad \Bigr{|} \quad }
{#1 \atop { }}}
\def\qsymmin#1 {{\quad \bigl {|} \quad \over \quad \bigr{|} \quad }
{#1 \atop { }}}

\def \overcross { { \atop / } {\hskip-5pt} { \biggl \backslash}
       {\hskip-5pt} {/ \atop } }
\def \cross {{\biggl {/ }} {\hskip-10pt} {\biggl {\backslash }}}

\line{\hfil DAMTP-R94/30}
\line{\hfil gr-qc/9408013}
\line{\hfil 10th August, 1994}
\medskip
\vfil

\centerline {\bf Spin networks, Turaev-Viro theory and the loop
representation.}
\line{ }

\centerline {\bf Timothy J. Foxon
{\footnote{${}^{\dagger}$}{\rm e-mail: tjf12@amtp.cam.ac.uk}} }
\smallskip
\centerline {\it Department of Applied Mathematics and Theoretical
Physics,}
\centerline {\it University of Cambridge,}
\centerline {\it Silver Street, Cambridge, CB3 9EW, U.K.}
 \line{ }

\vfil

\noindent {\bf Abstract.}
\smallskip

 We investigate the Ponzano-Regge and
Turaev-Viro topological field theories using spin networks and their
$q$-deformed analogues. I propose a new description of the state space for
the Turaev-Viro theory in terms of skein space, to which $q$-spin networks
belong, and give a similar description of the Ponzano-Regge state space
using spin networks.
I give a definition of the inner product on the skein space  and show
that this corresponds to the topological inner product, defined as the
manifold invariant for the union of two 3-manifolds.

Finally, we look at the relation with the loop representation of quantum
general relativity, due to Rovelli and Smolin, and suggest that the above
inner product may define an inner product on the loop state space.

\vfil
\eject

\beginsection {1. Introduction.}

 In this paper, we use the ideas of spin networks to examine the relation
between certain discrete models for gravity, based on topological field
theories, and the loop representation of quantum general relativity,
based on Ashtekar variables.

Spin networks were introduced by Penrose [1971] in an
attempt to use combinatorial ideas to construct a discrete model for a
quantum mechanical structure underlying classical space-time. This model
anticipates many of the ideas that were later formalised in topological
field theories.
 A spin network is an
$SU(2)$ invariant tensor represented by a trivalent graph. In such a
network, three labelled edges meet at each vertex, where each edge is
labelled by a half-integer representation $j_i$  of the
group $SU(2)$.

A 3-dimensional \tft is defined [Atiyah
1990] by the assignment of a topological invariant to a closed
3-manifold and a vector space to a 2-dimensional surface.  These
assignments are specified by defining the partition function of the
theory. We consider, in particular, topological field theories originally
defined on a triangulated 3-manifold by a partition function which is a
sum over states on the interior of the manifold. A state is defined
by labelling or `colouring' each edge of the triangulation by a half-integer
representation $j_i$ of the classical group $SU(2)$
[Ponzano and Regge 1968] or the quantum group $U_q(sl(2))$ at $q$ an
$r$-th root of unity [Turaev and Viro 1992].

 We begin, in section 2, by defining the state space and the inner product for
the Regge-Ponzano model and reviewing the previous work of Ooguri [1992].
In section 3, I show that there is a natural interpretation in terms of
 a spin network picture and extend the work of Rovelli [1993] to relate this
to the loop representation [Rovelli and Smolin 1990].
 In section 4, I describe a conjectured
new description of the state space of
the Turaev-Viro theory, obtained using the $q$-deformed version of spin
networks [Kauffman 1991], and re-write the inner product in this picture.
Finally, in section 5, we discuss the possible implications of these ideas, and
make the tenative proposal that the Turaev-Viro theory may be identified as a
naturally regularised version of the loop representation.

\beginsection {2. State space for Regge-Ponzano theory.}

 The \tft of Turaev and Viro [1992] is  defined by taking a triangulation of
a 3-manifold $M$, on which a state or colouring is given by labelling each
edge of the triangulation by a representation $j_i$ of the quantum group
$U_q(sl(2))$.
 The partition function $Z_{TV} (M)$ for the manifold $M$   is given by a
finite
sum over states on the interior of the manifold of the product of $q$-$6j$
symbols $\{6j^{(t)}\}_q$ corresponding to tetrahedra $t$, weighted by
specified functions $w_v(j_i)$ for each vertex,  $w_e(j_i)$ for each internal
edge, and $f(c_j)$ for the colouring $c_j$ on the boundary,
 $$ Z_{TV} (M, c_j) =
\sum_{states,\{ j_i \}} f(c_j) \, \prod_{vertices, v} w_v(j_i) \,
     \prod_{edges, e} w_e(j_i) \,
\prod_{tetra, t}
     \biggl\{
    \matrix  {j_1^{(t)} & j_2^{(t)} & j_3^{(t)} \cr
                    j_4^{(t)} & j_5^{(t)} & j_6^{(t)} \cr}
    \biggr\}_q, \eqno(2.1) $$
 for states which satisfy certain admissibility conditions, as we shall
discuss below. For a closed 3-manifold,
this partition function is a topological invariant, i.e. it depends only on
the topology of the manifold  $M$ and not on the triangulation used in the
definition. For a 3-manifold with boundary, the partition function is a
function of the colouring on the boundary, but still depends only on the
topology of the interior of the manifold.

  In the $q \to 1$ limit, this reduces
to the much earlier theory of Ponzano and Regge [1968], in which the $j_i$
label representations of the classical group $SU(2)$. The Regge-Ponzano
partition function $Z_{RP} (M)$ has a similar form to (2.1) but with
the $\{6j^{(t)}\}_q$ replaced by classical $6j$ symbols $\{6j^{(t)}\}$.
Specificly, for manifold $M$ with colouring $c_j$ on its boundary $F$,
$$ \eqalign{
 Z_{RP} (M, c_j) =
\sum_{states,\{ j_i \}} \,
    &\prod_{edges \in F,k} (-1)^{2 j_k} \sqrt {2 j_k +1} \,
     \prod_{vertices, v} \Lambda(j)^{-1} \cr
    &\prod_{edges,i} (2j_i + 1) \;
    \prod_{tetra, t}  (-1)^{\sum j_i} \,
     \biggl\{ \matrix  {j_1^{(t)} & j_2^{(t)} & j_3^{(t)} \cr
                    j_4^{(t)} & j_5^{(t)} & j_6^{(t)} \cr}
    \biggr\} . \cr
}\eqno(2.2) $$
 In this
case, the state sum is infinite, but it may be
regularised by truncating the sum at a large value $L$ of $j_i$, and the taking
the limit as $L \to \infty$. In this section, we shall assume that this may be
done, and interpret the Regge-Ponzano theory as a formal topological field
theory.

  To define fully  a 3-dimensional \tft [Atiyah 1990], the vector space of
states $V (F)$ assigned to a 2-dimensional surface $F$ must also be specified.
We define the vector space $V_{RP} (F)$ for Regge-Ponzano theory by
analogy with the  Turaev-Viro definition. This is essentially equivalent to
that given by  Ooguri [1992], but written in a way that allows us to give a
simple definition of the inner product.
 The vector space $V_{RP} (F)$  associated to a surface $F$, triangulated by
$\Delta$, is defined as a quotient of the space $C(F, \Delta)$ of linear
combinations of colourings of the triangulated surface, given by
$$\eqalign{
    V_{RP} (F) &=
   \Bigl\{ \phi_{\Delta} = \sum_i \lambda_i \, | \, c_i
\,\rangle\,
        :  {\rm P}[\phi_{\Delta}] = \phi_{\Delta} \Bigr\} \cr
          &= C(F, \Delta) \,
            \big{/}  \, \{ \phi_{\Delta}  :  {\rm P}[\phi_{\Delta}] =0 \, \},
         \cr
} \eqno(2.3)$$
where P is the projection operator corresponding to the
cylinder  $F \times [0,1]$ over the surface, given by
$$ {\rm P} | \, c_i \, \rangle \, = \,
     \sum_{c_i'} \, Z(F \times I, c_i, c_i')\; | \, c_i' \, \rangle \, .
\eqno(2.4)$$

  It is natural to assign the vector \ket M $\in V_{RP} (F)$ associated to a
3-manifold $M$ with boundary $ \partial M = F$  as
$$  | \, M \, \rangle = \sum_{c_i} \; Z_{RP}(M, c_i) \; | c_i \rangle \, ,
\eqno(2.5) $$
where $ c_i$ is a colouring of the boundary, and $Z_{RP}(M, c_i)$ is the
partition function given by a state sum of the form (2.2) over all colourings
of
the interior which extend the colouring $c_i$ on the boundary.

 An inner product may be defined on the space of colourings $C(F, \Delta)$  by
taking the set of different colourings on $F$ to form an orthonormal
basis,
$$ \braket{c_i \mid c_j} \, = \, \delta_{ij} .
    \eqno(2.6)$$
On the quotient space $V_{RP}(F)$,
we define the inner product  to be
$$ \inner {\xi_i}{\xi_j} = \braket{\xi_i \mid {\rm P}[ \xi_j]} ,
\eqno(2.7) $$
  so that,
for $\phi, \eta \in  \{ \phi_{\Delta}  :  {\rm P}[\phi_{\Delta}] =0 \, \}$,
as P is self-adjoint under (2.6),
we have
$$ \eqalignno{
      \inner {\xi_i + \phi}{\xi_j + \eta} &=
     \braket{\xi_i + \phi \mid {\rm P}[ \xi_j + \eta]}  \cr
 &= \braket{\xi_i \mid {\rm P}[ \xi_j]} \,
      + \, \braket{\phi \mid {\rm P}[ \xi_j]} \cr
    &= \braket{\xi_i \mid {\rm P}[ \xi_j]} \,
        + \, \overline{ \braket{\xi_j \mid  {\rm P}[ \phi]} } \cr
     &= \inner {\xi_i}{\xi_j} \, , &(2.8)\cr
}$$
and so the inner product is well-defined on  $V_{RP}(F)$.

 The above definitions imply that the
inner product between the two  vectors corresponding to two 3-manifolds
$M_1, M_2$, which meet in a common surface $F$,
 is given by the invariant of the manifold obtained by gluing $M_1$ and $M_2$
along $F$, as follows
$$\eqalignno{
    \inner {M_2}{M_1}
   &= \sum_{c_i, c_j, c_i' } \,
   \overline {Z(M_2, c_i)} \; Z(F \times I, c_j, c_i')\;  Z(M_1, c_j) \;
   \langle c_i \mid c_i' \rangle \cr
   &=  \sum_{c_i,c_j} \,
    \overline {Z(M_2, c_i)} \;  Z(F \times I, c_j, c_i)\; Z(M_1, c_j) \cr
   &= Z_{RP} (M_2 \cup M_1) \, . &(2.9) \cr
}$$
 This is the definition of the
inner product required for a  topological field theory [Atiyah 1990] to be
consistent, and so this definition carries a lot of information about the
elements of the theory.

 Witten [1989a]  defined a set of
topological field theories in which the partition function $Z_{CSW}(M)$
is given by a Chern-Simons path integral
$$  Z_{CSW}(M) = \int [{\d}A] \; \exp
  \biggl( i \; {k \over 4\pi}
  \int_M {\rm Tr } \, ( A \wedge {\d}A + {2 \over 3} A \wedge A \wedge A)
\biggr)  \eqno(2.10)
$$
 with the connection $A$ taking values
in the Lie algebras of various gauge groups. He argued
[Witten 1989b,c] that Chern-Simons theories with certain choices of gauge
group are equivalent to 2+1-dimensional gravity with or without
cosmological constant.
It follows that the Chern-Simons theory with gauge group $ISO(3)$ is
equivalent to 3-dimensional Euclidean gravity without
cosmological constant. This theory is related to the Regge-Ponzano model as
follows.

 The state space $V_{CSW}(F)$ of the
$ISO(3)$ theory consists of  functionals $\psi (\omega)$ on the moduli
space of flat $SO(3)$ connections on a 2-surface  $F$.
The wave function for a particular manifold $M$ is given by the
functional $\psi_M (\omega) \equiv \braket{\omega_i \mid \psi_M} $
defined by the path integral (2.10), with boundary condition that
$\omega_i$  is the flat connection on the boundary $\partial M = F$.
There is a natural inner product on this space
given by $$  \braket{\psi_{1} ( \omega) \mid \psi_{2}(\omega) }
     = \int_F [d\omega] \, \delta(F_{ij}) \,
      \overline{ \psi_{1}(\omega)}
\, \psi_{2}(\omega) .
\eqno(2.11)$$
It follows that the inner product of two Chern-Simons wave
functions corresponding to two handlebodies $M_1$ and $M_2$ is given by
the partition function for the closed, oriented manifold $M$
formed by gluing together the two handlebodies,
$$ \braket{\psi_{M_1}(\omega) \mid \psi_{M_2}(\omega) }
     = Z_{CSW} (M_2 \cup M_1) .
    \eqno(2.12) $$

   Ooguri [1992] related the two theories by constucting the
trivalent graph $W$, dual to the triangulation $\Delta$ on the
2-dimensional boundary $\partial M$ . Each
edge $C_i$ of this graph is labelled by the representation $j_i$ on the edge of
the triangulation that it crosses. This graph is interpreted as a Wilson line
network by assigning, to edge $C_i$, the Wilson line
$$ U_{j_i} [A, C_i] = {\rm P} \exp \;
\biggl \{
                           \int_{C_i} \; {A_a}^I \; {t_{j_i}}^J \; {\d} x^a
\biggr \} \, ,
    \eqno (2.13)
$$
where $t_{j_i}$ are the $SO(3)$ generators in the $j_i$ representation.
The product of these Wilson lines, contracted by $3j$ symbols at trivalent
vertices, gives a gauge-invariant function  $ \psi_{\Delta, c} (\omega)$
of the triangulation $\Delta$, the colouring $c$ and the flat connection
$\omega$. This may be interpreted as a change of basis function
$$ \braket{\omega \mid \Delta, c } \equiv \;
 \psi_{\Delta, c} (\omega), \eqno(2.14) $$
between a Regge-Ponzano state
$| \, \psi_M \rangle = \sum_c \mid {\Delta,c} \rangle \, \braket {\Delta,c
\mid \psi_M}  \in V_{RP}(F)$
and a Chern-Simons wave-function
$\braket {\omega \mid \psi_M} \equiv\psi_M (\omega) \in V_{CSW}(F)$,  as
$$ \braket  {\omega \mid \psi} =
    \sum_{c \in C(\Delta)}  \; \braket {\omega \mid \Delta, c } \,
                                              \braket {\Delta,c \mid \phi } ,
\eqno(2.15) $$
where the sum is over all colourings $c$ of the (fixed) triangulation
$\Delta$ on $F$.

  Ooguri showed that this correspondance is 1-1 and independent of the
triangulation chosen, and so the state spaces of the two theories
$V_{RP}(F)$ and $V_{CSW}(F)$ are isomorphic. He also claimed that the
two inner products given by (2.9) and (2.12) are equivalent, and so the
Regge-Ponzano theory and the $ISO(3)$ Chern-Simons theory are
equivalent as physical theories, (subject to the proviso that neither theory
is mathematically well-defined). This result suggests that  the
Regge-Ponzano theory may be interpreted as providing a model for quantum
gravity in three dimensions, and that we should look for a similar result
for the well-defined Turaev-Viro theory.

\beginsection { 3. Spin networks and the inner product.}

Spin networks were introduced by Penrose [1971] as a
discrete model underlying space-time.
 A spin network is an
$SU(2)$ invariant tensor represented by a trivalent graph,  a
network with three edges meeting at each vertex. Each edge is labelled
by a spin-$j$ representation of the angular momentum covering group
$SU(2)$,  where $j \in {0, {1\over 2},1, \ldots}.$
For the spin network to be non-zero, the triangle inequalities must hold at
 each vertex,
$$ | j_2 - j_3 | \leq j_1 \leq j_2 + j_3, \eqno(3.1)$$
and also, the sum of the spins must be integral,
$$ j_1 + j_2 +j_3 \in \hbox{\bf Z}, \eqno (3.2)$$
  where $ j_1, j_2, j_3 $ are the spins assigned to the three edges meeting
at that vertex.

  Penrose  showed how to evaluate a spin network
by using the binor
calculus [Penrose 1972].  In
this approach, a strand network is associated with each spin
network by replacing each edge, labelled by spin $ j $, by a linear combination
of $2j$ strands, and summing over all ways of joining these strands at
each vertex. So, a vertex is replaced in the strand picture by
$$ { {\; \; i \; {\biggl{\backslash} \biggl{/} \, j}} \atop \; \; {\bullet
\atop
  {\;\; \,\biggl {|} \, k }}   }
\quad = \quad
{  {2i \;  { \; \bigl{\backslash} \over \quad {\backslash  }} \; \,
{   \bigl{/}   \over \; {/  } \quad} \; 2j}
 \atop {\quad  {   { {\overline\bigvee } \atop |}  \over \quad
\bigl{|}\quad } { { } \atop 2k}}  }
 \eqno (3.3)$$
where each bar represents an anti-symmetriser, given by the linear
combination of the different ways the $2j$ strands may cross, with a
co-efficient of $(-1)$ for each crossing, e.g.
$$ \eqalign{
      \qsymmin 2 &= \; \bigg{|} \, \bigg{|} \; - \;
    \cross
\cr \cr
    \qsymmin 3    &= \; \bigg{|} \, \bigg{|} \, \bigg{|} \;
        - \; \bigg{|} \, \cross  \;
        - \; \cross  \, \bigg{|} \;
        + \; \cross {\hskip-7pt}
          \biggl{\backslash}  \;
        + \; { \biggl/ } {\hskip-7pt} \cross \;
        - \;  \cross  {\hskip-7pt} \bigg{|}
}\eqno(3.4)
$$

 An oriented abstract spin network (where the
orientation may be given by the embedding of the   \spnet in
the plane) is evaluated by decomposing each strand network into a number of
closed loops using the two basic binor identities
 [Penrose 1971, 1972],
$$\eqalignno{
   \cross \;  &=
   \; - \; {{\bigcup \atop \bigcap}} \;
   - \; {\biggr{)} \biggl{(} }  &(3.5)\cr \cr
  {\bigcirc} \;  &=  \; -2. &(3.6) \cr  }$$

These identities form the basis of a topologically invariant diagrammatic
calculus [Kauffman 1990], based on the SU(2) invariant tensor
$\varepsilon_{AB}$ represented by
$$\bigcap  = i \, \varepsilon_{AB} = i\, \pmatrix { 0&1\cr -1&0\cr}.
\eqno(3.7)$$
The binor calculus provides a way of calculating the norm of an abstract
oriented spin network, which only depends on the network as an abstract graph
together with its orientation. As we shall see in the next section, Kauffman
[1991] showed that the binor identity is a special case of his
bracket identity, which can be used to define a spin network based on the
quantum group  $sl(2)_q$.

We can now use spin networks to look at
the relation of the Regge-Ponzano model to the loop representation of
quantum general relativity [Rovelli and Smolin 1989], in a similar way to the
work of Rovelli [1993].
  Rovelli related Ooguri's Wilson line network to the loop representation
as follows.  Each Wilson line $U_{j_i} [A, C_i] $
in the $j$ representation is replaced by $2j$ Wilson lines in the ${1 \over
2}$ representation, and  each trivalent intersection is replaced by the sum
over all ways of joining the strands. The strand network
corresponding to a coloured triangulation  $(\Delta, c)$ is thus an ensemble
of multiple loops,  $E_{\Delta, c} = \{\alpha_1, \alpha_2, \ldots \}, $ where
each multiple loop $\alpha_i $ has the property that $2j$ single loops
cross a link of the triangulation with colour $j$.

   Here, we interpret the dual graph to a coloured triangulation in the
Regge-Ponzano model as a spin network.
This was first suggested by Hasslacher and Perry [1981] and Moussouris
[1983] for the case of a spin network on $S^2$. This spin network is converted
to a strand network by replacing each trivalent vertex by a sum over
strands according to (3.3). This strand network is then evaluated using
the binor relations (3.5),(3.6). The key difference here from Rovelli's
approach is that the presence of the extra negative signs in the
anti-symmetrisers (3.4). This rule, together with the binor relations, ensures
that the evaluation of the spin network is topologically invariant. We shall
see
in the next section that this is because the binor relations (3.5),(3.6) are a
special case of the Kauffman relations.

 Now, the state space of the Regge-Ponzano model $V_{RP} (F)$ is given by
the space of linear combinations of colourings of a triangulation. We map a
particular coloured triangulation to its dual spin network, by mapping a
coloured edge of the triangulation to an edge of the network crossing it with
the same colouring, as follows
$$ i { { } \atop {  \Biggl{/}  \Biggl{\backslash} \over { k} }  } j
     \quad \longrightarrow \quad
    { {\; \; i \; {\biggl{\backslash} \biggl{/} \, j} } \atop \; \; {\bullet
\atop
  {\; \; \, \biggl {|} \, k }}   }
\eqno (3.8)$$
 It is clear that this mapping is  1-1. A spin
network is evaluated as a linear combination of strand networks which obey
the binor identities and so are isotopy invariant. This suggests that we can
re-write the state space $V_{RP} (F)$ as the space of linear combinations of
isotopy classes of links, quotiented by the binor relations. As we shall see in
the next section, this is an example of a skein space.

  A basis element in the Regge-Ponzano state space $V_{RP} (F)$  given  by a
paticular colouring of a triangulation \ket {\Delta, c_i}
maps to a spin network state \ket s_i . A spin network state corresponds to a
 linear combination of strand network states \ket \alpha_i as above,
 $$ | \Delta, c_i \rangle \to
\quad | s_i \rangle \; =
    \; \sum_{\alpha_i \in s_i} \, (-1)^{n} \, | \alpha_i \rangle ,
\eqno(3.9)$$
where $n$ is the number of crossings in that state.
  We now map a strand network state to an equivalence class of a  multiple
loop under the binor relations (3.5),(3.6). So, a spin network
\ket s_i $\in V_{RP} (F)$  maps to a state
in the space $V_{loop} (F)$  of complex linear combinations of characteristic
functionals on isotopy classes of loops, quotiented by the binor relations.

   We now want to consider the question of how to define an inner product
on the state space. Rovelli proposed that mappings of the above type could be
used to define the inner product between two loop states using Regge-Ponzano
theory.
 From the above
formulation, we propose the following interpretation. As a coloured
triangulation maps 1-1 on to a spin network, this suggests that, following
(2.6), we should take  spin network states with different colourings to be
orthonormal, $$ \braket {s_i \mid s_j} = \delta_{ij}.
    \eqno(3.10)$$

 An arbitrary loop state
\ket \gamma_i $\in V_{loop} (F)$ may be written as
a linear combination of spin network states \ket s_j as
$$ \langle \gamma_i | \; = \; \sum_j \, \lambda_j \, \langle  s_j |.
\eqno(3.11)$$
For a particular state corresponding to the vector for a 3-manifold,
\ket \gamma_M , the co-efficient is $\lambda_j = Z(M, s_j) $, and so
$$\eqalignno{
 \braket {\gamma_{M_1} \mid \gamma_{M_2}} &=
     Z_{RP} (M_1 \cup M_2). & (3.12) \cr
}$$

  We shall discuss in section 5, the comparison between this approach and
that of Rovelli and Smolin [1990].

\beginsection {4. Turaev-Viro theory and skein space.}

   I now want to extend the above work to begin to investigate the relation
between the topological field theory of Turaev and Viro [1992] and the
loop representation of quantum general relativity in 3 dimensions. There
are two pieces of evidence which lead us to suspect such a relation. Firstly,
Turaev-Viro theory is the generalisation of the Regge-Ponzano theory in
which the representations of the classical group $SU(2)$ are replaced by
those of a quantum group $U_q(sl(2))$ at $q$ an $r$-th root of unity, so
that the partition function is a finite sum. Thus, we may expect that the
Turaev-Viro theory would be related to the loop representation. Secondly, we
know that the Turaev-Viro invariant is the square of the Chern-Simons-Witten
invariant for gauge group $SU(2)$, which is equivalent to the $SO(4)$
Chern-Simons theory that is related to 3-dimensional Euclidean gravity
with a positive cosmological constant. This formal relation was implicit in
Witten's work and was first written explicitly in [Ooguri and Sasakura
1991] and [Archer and Williams 1991], as follows
$$ \eqalignno{
 Z_{TV,r} (M)  &= {\bigl | Z_{CSW,k} (M) \bigr |} ^2    \cr
&= \int [{\d}A, {\d}B] \; \exp
   \biggl( i \; {k \over 4\pi} \int ( A \, {\d}A + {2 \over 3} A^3) \, - \,
     ( B \, {\d}B + {2 \over 3} B^3) \biggr) \cr
&= \int [{\d}e, {\d}\omega] \; \exp
   \biggl( i \int ( e \wedge R \, + \,
{\Lambda_k \over 3} \, e \wedge e \wedge e) \biggr), &(4.1) \cr
}$$
where
$$ \eqalign {
 \omega &= {1\over 2} \, (A+B), \quad
  e = {k \over 8\pi} \, (A - B),   \cr
  \Lambda_k &= \Bigl({4\pi \over k} \Bigr)^2 , \cr
} \eqno(4.2)$$
for connections $A,B$ taking values in the Lie algebra of $SU(2)$,
connection $\omega$ taking values in $SO(3)$, triad $e$ and cosmological
constant $\Lambda_k$, for level $k$ of the Chern-Simons theory, which is
related to the level $r$ of the Turaev-Viro theory by
$$ k = r - 2. \eqno(4.3) $$

  Much work has been undertaken by mathematicians to provide a more
mathematically rigorous proof of the Chern-Simons-Witten topological
field theory with gauge group $SU(2)$ which does not rely on the use of
path integrals. Reshetikhin and Turaev  [1991] defined a topological
invariant based on representations of the quantum group $U_q(sl(2))$
which satisfies the same formal properties and so may be identified with the
Chern-Simons-Witten invariant.
In a series of papers, Lickorish [1991,1993] showed  that this
Witten-Reshetikhin-Turaev (W-R-T) invariant could be reproduced using
skein theory. Independently, both Turaev [1992a,b] and Walker [1992]
showed that the Turaev-Viro invariant is the square of the W-R-T
invariant, and so justified equation (4.1).
More insight into the relation between these two theories was
provided by the work of Justin Roberts [Roberts 1993,1994] extending the
skein-theoretic approach, and here we shall extend this approach.

In a physical model, we are interested in finding an interpretation
of the vector space of states and the inner product between states defined by
a topological field theory. From the relation (4.1) between the partition
functions and the formalism of  a \tft [Atiyah 1990], it is expected that the
state space for the Turaev-Viro theory to be isomorphic to endomorphisms of
the state space for the Witten theory, $$ V_{TV} (F) \cong {\rm End }
\;V_{WRT} (F) .
     \eqno(4.4) $$
An outline proof of this theorem was given in [Turaev 1992b]. We
show how this relation is recovered in the skein-theoretic approach.

 The skein space
{\sk M} of an oriented 3-manifold $M$ is the vector space of formal
linear sums, over {\bf C} , of isotopy classes of framed links $L$ in
$M$, quotiented by the Kauffman relations [Kauffman 1991]
$$\eqalignno{
  \overcross
   &= A \; {{\bigcup \atop \bigcap} } \;
      + \; A^{-1} \; {\Bigl{)} \Bigl{(} } &(4.5) \cr \cr
   {\bigcirc} \; \cup \; L  &= \; ( - A^2 - A^{-2}) \; L, &(4.6) \cr
}$$
where  $\bigcirc$ is a closed contractible loop, and the
diagrams in (4.5) represent pieces of links inside a 3-ball, outside of which
the links are identical. Applying (4.5) to a link reduces that link to a linear
combination of loops with no crossings inside any local 3-ball, and each
contractible loop may be replaced by a co-efficient by (4.6). Thus, the
content of the skein space is that it is the space of linear combinations of
closed loops with no local crossings which are both non-contractible and
disjoint.

 We can see that the Penrose binor relations (3.5),(3.6) are a special of the
Kauffman skein relations, given by taking $A = -1$.

  It is convenient to use a two-dimensional projection to describe the
skein space, but we can see that the above definition is inherently
three-dimensional. In particular, the skein space ${\cal S} A$  of a solid
torus $S^2 \times S^1$ is described by its projection on to a
2-dimensional annulus $A$ and is given by the polynomial algebra over
$\alpha_n$, where $\alpha_n$ is the basis element consisting of $n$
loops around the hole.
We can also use this  description as a projection to define the skein space
\sk{F} of a 2-surface $F$ as the skein space of the cylinder
$F \times [0,1]$ over the surface.

  A description of the state space of a surface in the W-R-T theory using
skein theory was given by Roberts  [1994] as follows.  Consider the surface
$F$ as dividing $S^3$ into a handlebody $H$ and its dual $H'$,
$$ S^3 = H \, \cup \, (F \times I) \, \cup H' ,
\eqno(4.7)$$
and define the bilinear form
$$ \braket{\; , \,} : {\cal S} H \times {\cal S} H' \to {\cal S} S^3 =
    \hbox{\bf C}, \eqno(4.8)$$
then, on the quotient ${\cal H}$ of the skein space  of the handlebody
defined by
 $$ \eqalignno{
U &= {\rm Ker } \, ( {\cal S} H \to ({\cal S} H')^*)
& (4.9) \cr
{\cal H} &= {\cal S} H / U & (4.10)
}$$
for $q$ a $2r$-th root of unity, the form descends to a non-degenerate
pairing
$$ {\cal H} \times {\cal H'} \to \hbox{\bf C}.
\eqno(4.11)$$
 This implies [Blanchet et al 1993]
that the space ${\cal H}$ is finite dimensional and may be
identified as the state space of the W-R-T theory
$$ V_{WRT} (F) \cong  {\cal H}, \quad (F = \partial H).
\eqno(4.12)$$

  We now consider the $q$-deformed analogues of spin networks
[Kauffman 1991,1992].
As in the undeformed `classical' case, a $q$-spin network is a labelled
trivalent graph, but now the labellings refer to integer representations of the
quantum group $sl(2)_q$. Again, a trivalent vertex or triad corresponds to a
linear combination of connected strands, but the anti-symmetriser is now
replaced by a $q$-symmetriser or idempotent $f^{(j)}$. This represents a
linear combination of braided strands, i.e.  one strand locally overcrosses
another but the strands do not intersect.
So, a triad $T(i,j,k)$ is an element of the
skein space of the form

$$ { {\; \; i \; {\biggl{\backslash} \biggl{/} \, j}} \atop \; \; {\bullet
\atop
  {\;\; \,\biggl {|} \, k }}   }
\quad = \quad
{ \;  i \;  { \; \bigl{\backslash} \over \quad {\backslash  }} \; \,
{   \bigl{/}   \over \; {/  } \quad} \; j
 \atop \quad  {   { {\overline\bigvee } \atop |}  \over \quad
\bigl{|}\quad } { { } \atop k}}
 \eqno (4.13)$$
where a $q$-symmetriser is given by the linear
combination of the different ways of braiding the $2j$ strands with a
co-efficient of $(+A^{-3})$ for each crossing, e.g.
$$ \eqalign{
      \qsymmin 2 &= \; \bigg{|} \, \bigg{|} \, + \, A^{-3} \,
    \overcross
\cr \cr
    \qsymmin 3    &= \; \bigg{|} \, \bigg{|} \, \bigg{|} \,
        + \, A^{-3} \, \bigg{|} \, \overcross  \,
        + \, A^{-3} \, \overcross  \, \bigg{|} \,
        + \, A^{-6} \, { { \atop / } {\hskip-5pt} { \biggl \backslash}
{\hskip-5pt}
          \biggl{\backslash}  {\hskip-5pt} {/ \atop } }  \,
        + \, A^{-6} \, { \atop / } {\hskip-4pt}
          { { \atop / } {\hskip-6pt} { \biggl \backslash}
       {\hskip-6pt} {/ \atop } } {\hskip-4pt} {/ \atop } \,
        + \, A^{-9} \,  { { \atop / } {\hskip-5pt} { \biggl \backslash}
       {\hskip-5pt} {/ \atop } } {\hskip-10pt} {| \atop |}
}\eqno(4.14)
$$
In the limit $A \to -1$, which corresponds to $q = A^2 \to 1$, these reduce to
the anti-symmetrisers (3.4), and  so a $q$-spin network reduces to a classical
spin network as $q \to 1$.

  A $q$-spin network is evaluated by
applying the Kauffman relations  (4.5), (4.6) to each crossing and
contractible loop, and so is naturally an element of the skein space
{\sk M}. A triad $T(i,j,k)$ is defined for
$$\eqalign{
   i ,j,k \in \{ 0, 1, \ldots,  r-1 \} &,  \cr
    i \leq j+k,\; j \leq i+k, \; k \leq i+j &, \quad
  {1 \over 2} ( i +j +k) \in  \hbox{\bf Z}, \cr
 } \eqno(4.15)$$
and is said to be admissble if
$$\eqalign{
     i,j,k &\leq r-2,  \cr
   {1 \over 2} (i+j+k) &\leq (r-2),  \cr
}\eqno(4.16)$$
and inadmissible if these relations are not satisfied.
 Note that networks
are labelled by integers, whereas triangles in the Turaev-Viro picture
were labelled by half-integers.

  A $q$-spin network is admissible if and only if all its triads are
admissible.
As we shall see, we want to impose the condition that we only consider
admissible networks.
So, we define $N$ as the subspace of the skein space \sk{M} of any
3-manifold generated by inserting an inadmissible triad anywhere
into the manifold $M$, i.e. the subspace of isotopy classes of  $q$-spin
networks in $M$ for which any triad is inadmissible, and quotient out by this
subpsace.
 Roberts [1994] showed that this quotient space of the skein space of a
handlebody, $$ {\cal H} = {{{\cal S}H}/ N},
 \eqno(4.17)$$
which we call the reduced skein space,
is equivalent to the quotient space (4.10). So, the state
space for the Witten theory $V_{WRT} (F) \cong {\cal H}$ has now been
written as the quotient space of the skein space of the handlebody by the
subspace generated by inadmissible triads. We can regard this as the space of
admissible $q$-spin networks on the handlebody.

 Furthermore, Roberts showed that the skein space {\sk F} of the
2-surface  $F$ maps surjectively onto the endomorphisms of ${\cal H}$,
and so the quotient ${\cal F}$ of the skein space by the kernel of this map
$$ \eqalign{
   {\qquad Q} &= {\rm Ker } \, ( {\cal S} F \to ({\rm End} \, {\cal H} ))
 \cr
{\cal F} &= {\cal S} F / Q ,  \cr
} \eqno(4.18)$$
is isomorphic to the algebra of endomorphisms of ${\cal H}$,
$$ {\cal F} \cong {\rm End} \, {\cal H}. \eqno(4.19)
$$

  Comparing this relation with the expected relation (4.4) between the
state spaces of the two theories, I conjecture
that space ${\cal F}$ is isomorphic to the state space of the
Turaev-Viro theory $V_{TV} (F)$, and further that, in analogy with  Roberts'
result for the Witten state space, this space ${\cal F}$ is equivalent to the
reduced skein space defined by
 $$ {\cal F} = {\cal S} F / N,  \eqno(4.20)$$
where $N$  is the subspace of \sk{F} generated by inserting any
inadmissible triad into $F$.

  This relation may be expected from general principles.
Recall that the Turaev-Viro theory [Turaev and Viro 1992] is specified by the
partition function (2.1), which is given by a state sum over admissible
colourings of a triangulated 3-manifold.
The labels $(i,j,k)$
assigned to the three edges around any triangle must satisfy
$$\eqalign{
    (i,j,k)  \in \{ 0, {1\over 2},  1, \ldots, {(r - 2) \over 2} \} &,\cr
    i \leq j+k,\; j \leq i+k, \; k \leq i+j &,  \quad
    i +j +k \in \hbox{\bf Z},
}\eqno(4.21)$$
and the admissibility condition
$$  i+j+k \leq (r-2),  \eqno(4.22)
 $$
where the theory is defined at $q$ an $r$-th root of unity.

The state space $V_{TV} (F)$ of the Turaev-Viro theory
 is
defined as the space of linear combinations of admissible colourings $C (F)$
on the triangulated surface $F$, quotiented by the kernel of the isomorphism
$\Phi_{(F \times I)}$ corresponding to the cylinder $F \times [0,1]$ over
the surface,
$$ V_{TV} (F) = C (F) \, / \, {\rm Ker} \, \Phi_{(F \times I)}.
\eqno(4.23)$$

   An admissible colouring of a triangulation of the surface,
 \ket {\Delta, c} $\in  V_{TV} (F)$,
 may be mapped to a $q$-spin network, as follows.
Take the trivalent graph on the surface dual to the triangulation, as in (3.8),
and label an edge of the graph by twice the half-integer label on the edge of
the triangulation which it crosses. This integer-labelled trivalent graph may
now be interpreted as a $q$-spin network.
Furthermore, a colouring
of a triangulation which is admissible according to (4.21), (4.22)  maps
to a network in which the colouring of each triad is admissible by
(4.16). Thus,  the map from the Turaev-Viro state space
$V_{TV} (F)$ to the reduced skein space ${\cal F}$ given by
$$ \phi : V_{TV} (F) \to
{\cal F} \eqno(4.24)$$
is  surjective and we expect it to be an isomorphism.

 To gain more physical insight into the nature of the reduced skein space
${\cal F}$,
we may attempt to find a basis for this space.  A basis is not given simply by
the set of all admissible colourings of a $q$-spin network dual to a simple
triangulation of the surface, as there is a subtlety due to the fact that
this space is the skein space quotiented by the space generated by
inadmissible triads.

  Let us consider the simple example in which $F$ is the surface of the
torus $S^1 \times S^1$ and $r=4$ so that the admissible colourings of an
edge are $j \in \{0,1,2\}$. The Witten state space for this surface is the
reduced skein space ${\cal H}$ of the solid torus, which has a basis
 $\alpha_j$ given by $j\in \{0,1,2\}$ loops around the hole,  (since the
inadmissibility of triads of higher order imply that the skein element for
any higher number of loops must be zero),
  $$ {\cal H} = \Bigl\{ \sum_j \lambda_j \, \alpha_j \;
   : \lambda_j \in \hbox{\bf C}, \; j \in \{0,1,2\}  \Bigr\}
\eqno(4.25)$$
 and so is 3-dimensional. By the relation (4.4), the Turaev-Viro state space
$V_{TV} (F)$ should be 9-dimensional. The simplest (degenerate)
triangulation of the surface consists of two triangles with their edges
identified and so the dual $q$-spin network has two vertices and three
edges, labelled by $j_1, j_2, j_3$, such that $(j_1, j_2, j_3)$ form an
admissible triad. In this case, the admissible colourings of a triad are
$\{(0,0,0), (0,1,1),  (0,2,2), (1,1,2)\}$ and so, allowing for permutations,
there are 10 admissible colourings of this network. However,  these
colourings are not independent in the reduced skein space ${\cal F}$
because of  relations generated by inadmissible colourings. In particular,
we find that there is exactly one relation which may be generated by the
inadmissible colouring $(2,2,2)$ of this network, and so the space
${\cal F}$ is indeed 9-dimensional. Similarly, we find that, for the case
$r=5$, there are four relations amongst the 20 admissible colourings of
this network and so ${\cal F}$ is 16-dimensional. These examples
lend support to our conjecture that ${\cal F}$ may be identified as the
Turaev-Viro state space $V_{TV} (F)$.

  Comparing with the definition (2.3) for the Regge-Ponzano theory,
this suggests that
 we could define an inner product on the reduced skein space by taking
different colourings of a $q$-spin network to be orthonormal. This picture of
the state space should also enable us to relate a $q$-spin network on the
boundary to the
 the Kauffman-Lins
[1990] formulation of the Turaev-Viro theory in the interior of the 3-manifold.

\beginsection {5. Discussion.}

  In this paper, we have explored several aspects of the relation between spin
networks, simplicial state-sum models and the loop representation for
quantum general relativity. We conclude by summarising what we have learned
and discussing the possible implications.

  We have given descriptions of the state space of a surface for the
Ponzano-Regge theory (2.3) as the space of spin networks on that surface, and
for  the Turaev-Viro theory as the space of admissible $q$-spin networks on
the surface (4.24). The latter space is called the reduced skein space of the
surface.  The advantage of working with skein space is that isotopy
invariance is automatically encoded in the formalism. Indeed, the Kauffman
skein relations underly the Jones polynomial isotopy invariant (see [Kauffman
1991]). The other advantage is that the reduced skein space is isomorphic to
endomorphisms the state space for the Witten theory based on $SU(2)$, as
expected from the fact that the Turaev-Viro theory is the square of that
theory.

   We also considered the definition of the inner product for the
Regge-Ponzano theory.
   In any topological field theory, there corresponds a vector, in the state
space  for a surface,  to a manifold with that surface as its boundary.
The inner
product defined on the state space must be such that the inner product
of the two vectors corresponding to two manifolds which meet in a
common surface is given by the invariant of the manifold given by their
union. This guarantees that the invariant of a closed manifold does not
depend on how the manifold is cut into two pieces (see [Atiyah 1990]),
and so this inner product carries a great deal of information about the
elements of the theory. We saw that, for the Regge-Ponzano theory, we
could recover this topological inner product  from the inner product
defined by taking different colourings of a particular spin network on
the surface to be orthonormal.

 Building on the work of Ooguri [1992] and Rovelli [1993],  we considered the
relation between the Regge-Ponzano theory and the loop representation for
quantum general relativity.
  As emphasized by Ashtekar [1991], the problem of finding an inner product
on the space of states is one of the key problems in non-perturbative canonical
gravity.   Rovelli and Smolin [1990] defined the state space for quantum
general relativity as a quotient of the space of functionals on multiple loops.
Such loop states were related by the $SU(2)$ spinor relations,
from which the binor relations differ in the sign of the terms. They showed
that states given by functionals on link isotopy classes of simple,
non-intersecting loops satisfied the constraint equations, and so could be
identified as physical states. They defined an inner product by
taking characteristic functionals on isotopy classes of simple loops to be
orthonormal, (see  Smolin [1992]).

  Here, we took the loop state space to be the space of complex linear
combinations of isotopy classes of loops, and wrote an arbitrary loop state as
a linear combination of spin network states. We defined the inner product by
taking different colourings of a spin network to be orthonormal. The
advantages of this approach are that isotopy invariance is automatically
encoded by the binor relations which underly spin networks, and that the
inner product is consistent with the topological inner product, as explained
above. However, it is not clear whether suitable self-adjoint operators may
be defined under this inner product,
and so this interpretation must be regarded as
provisional.

   The above considerations, together with the work of Ooguri [1992], suggest
that the Regge-Ponzano theory may be regarded as a discrete model for
quantum gravity in three dimensions. Unfortunately, this theory is not
mathematically well-defined as the state sum (2.2) is infinite and it must be
regularised. Ooguri followed the regularisation of Ponzano and Regge, but it
is not clear how rigorous this is. We have ignored these problems  here and
treated the Regge-Ponzano theory as a formal topological field theory.
It is actually a generalisation of such a theory in the sense that the state
space $V_{RP}(F)$ is infinite-dimensional. Rovelli [1993] suggested that
quantum general relativity in four dimensions may be a generalised
topological field theory of this kind.

  However, at least in three dimensions, these problems with regularisation
may be completely overcome by considering instead the well-defined
Turaev-Viro theory. The state sum (2.1) is finite as there are only a finite
set
of edge labels (4.21) and so of possible colourings, for the theory defined at
$q$ an $r$-th root of unity. As $q \to 1$ and $r \to \infty$, the Turaev-Viro
theory reduces to the Ponzano-Regge theory and so it may be regarded as a
naturally regularised version of that theory. The Turaev-Viro state space
$V_{TV}(F)$ is finite-dimensional, and we have conjectured that it is
isomorphic to the reduced skein space of the surface $F$.
 As $r \to \infty$, the reduced skein space goes to the space of linear
combinations of isotopy classes of links quotiented by the binor relations,
which we identified as the Regge-Ponzano state space.

  This suggests that we may interpret the Turaev-Viro theory as a finite
discrete model for quantum gravity. As with the Ponzano-Regge theory,
we may define the inner product by taking different colourings of a
$q$-spin network to be orthonormal.
We would then need to check that this definition was equivalent to the
topological inner product.
If so, it would be interesting  to consider whether the Turaev-Viro theory
could be related to the  loop representation.

\beginsection {Acknowledgements.}

  I would like to thank John Barrett, Chris Fewster and Justin Roberts for
helpful comments and discussions.

\beginsection {References.}

\noindent  F.J. Archer and R.M. Williams 1991,
``The  Turaev-Viro state-sum
model and

3-dimensional quantum gravity,"
{\it Phys. Lett.} {\bf B273} (1991), 438-450.
\smallskip

\noindent A. Ashtekar 1991, {\it Non-perturbative canonical
gravity,} lecture notes
prepared

with R.S. Tate, (World Scientific, Singapore, 1991).
\smallskip

\noindent M. Atiyah 1990, {\it The geometry and physics of
knots,}
(Cambridge University Press, 1990).
\smallskip

\noindent C. Blanchet, N. Habegger, G. Masbaum and P. Vogel 1993,
``Topological quantum field

theories
derived from the Kauffman bracket,"
preprint, {\it University of Nantes,} 93-02/1.
\smallskip

\noindent  B. Hasslacher and M.J. Perry 1981, ``Spin
networks are simplicial
quantum gravity,"

{\it Phys. Lett.} {\bf 103B} (1981), 21-24.
\smallskip

\noindent L.H. Kauffman 1990, ``Spin networks and knot
polynomials,"

{\it Int. J. Mod. Phys.} {\bf A5} (1990), 93-115.
\smallskip

\noindent  L.H. Kauffman 1991, {\it Knots and physics,}
(World Scientific, Singapore, 1991).
\smallskip

\noindent  L.H. Kauffman 1992, ``Map colouring, $q$-deformed
spin networks, and
Turaev-Viro

invariants for 3-manifolds,"
{\it Int. J. Mod. Phys.} {\bf B6} (1992), 1765-1794.
\smallskip

\noindent  L.H. Kauffman and S. Lins 1990,
``A 3-manifold invariant by state
summation,"

preprint, {\it University of Illinois at Chicago,} (1990).
\smallskip

\noindent W.B.R. Lickorish 1991, ``Three-manifolds and the
Temperley-Lieb algebra,"

{\it Math. Ann.} {\bf 290} (1991), 657-670.
\smallskip

\noindent W.B.R. Lickorish 1993, ``The skein method for
three-manifold invariants,"

{\it  J. Knot Theory and its Ramifications} {\bf 2} (1993), 171-194.
\smallskip

\noindent  J.P. Moussouris 1983, ``Quantum models of
space-time based on recoupling
theory",

{\it D.Phil.} Thesis, (Oxford University, 1983).
\smallskip

\noindent H. Ooguri 1992, ``Partition functions and
topology-changing amplitudes in the 3D

lattice gravity of Ponzano and Regge," {\it Nucl. Phys.} {\bf B382} (1992),
276-304.
\smallskip

\noindent H. Ooguri and N. Sasakura 1991, ``Discrete
and continuum
approaches to three-dimensional

quantum gravity,"
{\it Mod. Phys. Lett.} {\bf A6} (1991), 3591-3600.
\smallskip

\noindent  R. Penrose 1971, ``Angular momentum: An approach to
combinatorial space-time,"

in {\it Quantum theory and beyond,} ed. T.Bastin, (C.U.P.,
Cambridge, 1971).
\smallskip

\noindent  R. Penrose 1972, ``Applications of negative
dimensional tensors,"
in {\it Combinatorial theory

and applications,} ed. D. Welsh, (Wiley, 1972).
\smallskip

\noindent  G. Ponzano and T. Regge 1968,
``Semiclassical limit of Racah
coefficients,"
in {\it Spectroscopic
and group theoretical methods in physics,} ed. F. Bloch,
(North Holland, 1968).
\smallskip

\noindent  N.Y. Reshetikhin and V.G. Turaev 1991, ``Invariants of 3-manifolds
via link polynomials

and quantum groups,"
{\it Invent. Math.} {\bf 103} (1991), 547-597.
\smallskip

\noindent J.D. Roberts 1993, ``Skein theory and Turaev-Viro
invariants,"

preprint, {\it DPMMS, Cambridge.}
\smallskip

\noindent  J.D. Roberts 1994, ``Skeins and mapping class groups,"

{\it Math. Proc. Camb. Phil. Soc.} {\bf 115} (1994), 53-77.
\smallskip

\noindent C. Rovelli 1993, ``Basis of the
Ponzano-Regge-Turaev-Viro-Ooguri quantum
gravity model

is the loop
representation basis,"
{\it Phys. Rev.} {\bf D 48} (1993), 2702-2707.
\smallskip

\noindent C. Rovelli and L. Smolin 1990, ``Loop
representation for quantum
general relativity,"

{\it Nucl. Phys.} {\bf B331} (1990), 80-152.
\smallskip

\noindent   L. Smolin 1991, ``Recent developments in
non-perturbative quantum gravity,"

in {\it Proceedings of the 1991 GIFT International Seminar on Quantum
Gravity and

Cosmology,}
(World Scientific, Singapore, {\it in press}).
\smallskip

\noindent V.G. Turaev 1992a, ``Quantum invariants of 3-manifolds and a
glimpse of shadow

topology,"
{\it Lect. Notes in Math.} {\bf 1510}, 363-366.
\smallskip

\noindent V.G. Turaev 1992b, ``Quantum invariants of 3-manifolds,"

preprint, {\it Institut de Recherche Math\'ematique Avanc\'ee, Strasbourg.}
\smallskip

\noindent V.G. Turaev and O.Y. Viro 1990, `` State sum
invariants of 3-manifolds
and quantum

$6j$ symbols," {\it Topology } {\bf 31} (1992), 865-902.
\smallskip

\noindent K. Walker 1992, ``On Witten's 3-manifold invariants,"
preprint.
\smallskip

\noindent E. Witten 1989a, ``Quantum field theory and the
Jones polynomial,"

{\it Commun. Math. Phys.} {\bf 121} (1989), 351-399.
\smallskip

\noindent E. Witten 1989b, ``2+1 dimensional gravity as a
exactly soluble theory,"

{\it Nucl. Phys.} {\bf B311} (1989), 46-78.
\smallskip

\noindent  E. Witten 1989c, ``Topology-changing amplitudes in
2+1 dimensional gravity,"

{\it Nucl. Phys.} {\bf B323} (1989), 113-140.
\smallskip

\bye